
\input harvmac
\input epsf
\noblackbox

\Title {\vbox{\baselineskip12pt\hbox{BUHEP-94-37}
\hbox{hep-ph@xxx/9412309}}}
{\vbox{\centerline{Limits on the Ununified Standard Model}\vskip2pt}}


\centerline{R.S. Chivukula, E.H. Simmons and J. Terning\footnote{}
{e-mail addresses: sekhar@abel.bu.edu, simmons@smyrd.bu.edu,
terning@calvin.bu.edu}}
\bigskip
\centerline{Department of Physics}
\centerline{Boston University}
\centerline{590 Commonwealth Ave.}
\centerline{Boston, MA 02215}


\vskip .3in

The ununified standard model is an extension of the standard model
that contains separate electroweak gauge groups for quarks and leptons.
When it was originally proposed, data allowed the new gauge bosons
 to be quite light. We use recent data from precision electroweak measurements
to put stringent bounds on the ununified standard model. In particular, at the
95\% confidence level, we find that the
ununified gauge bosons must have masses above about 2 TeV.
\Date{Dec. 15, 1994} 



\def\l{\ell}
\def\st{\sin\theta_W}
\def\ct{\cos\theta_W}
\def\sp{\sin\phi}
\def\cp{\cos\phi}

\def\gp{g^\prime}
\def\gq{g_q}
\def\gl{g_\ell}

\def\gtap{\raise.3ex\hbox{$>$\kern-.75em\lower1ex\hbox{$\sim$}}}


\def\su2{SU(2)}

\newsec{Introduction}

The standard $SU(2) \times U(1)$ electroweak model is in satisfactory
agreement with the panoply of data from precision measurements
at LEP and SLC as well as low-energy experiments. Numerous extensions
to this model predict the existence of extended weak gauge-symmetries
and, hence, of additional weak-charged gauge-bosons. However, even
prior to the advent of precision measurements at LEP and SLC, the
extra gauge-bosons in most such models were constrained to be heavy
because of their potential effects on low-energy measurements and on
the $W$ and $Z$ masses. One interesting counter-example to this is the
ununified standard model \ref\oldunun{H. Georgi, E.E. Jenkins and E.H.
Simmons, {\it Phys. Rev. Lett} {\bf 62} (1989) 2789 and {\it Nucl.
Phys.} {\bf B331} (1990) 541.} which contains separate electroweak gauge
groups for quarks and leptons. To a good approximation in this theory
the existence of extra weak-charged gauge-bosons {\it does not} spoil
the tree-level relationship between $G_F$ as measured in $\mu$-decay,
$\sin^2\theta_W$ as measured in deep-inelastic $\nu$-scattering, and
$M_W$ or $M_Z$.  Hence, those data could accommodate extra states as
light as 250 GeV \oldunun .

At the time that the ununified model was proposed, it was anticipated
that the model would be more stringently tested by high-energy data
such as measurements of $Z$ branching ratios at LEP \oldunun\ and
measurements of forward-backward asymmetries at LEP and the Tevatron
\ref\unrandall{L. Randall, {\it Phys. Lett.} {\bf B234} (1990)
508.}. Indeed, as high-energy data from LEP and HERA became available,
the lower bound on the masses of the extra gauge bosons was raised to
roughly 500 GeV (the precise value depending on the strength of the
mixing between the sets of weak gauge bosons) \ref\rizzos{T.G. Rizzo,
{\it Int. J. Mod. Phys.}  {\bf A7} (1992) 91; V. Barger and T. Rizzo,
{\it Phys. Rev.} {\bf D41} (1990) 946.}.   Later, Tevatron dijet data
was shown to give similar limits \ref\rizzotev{T.G. Rizzo, hep-ph/9302253,
{\it Phys. Rev.} {\bf D48} (1993) 4470.}.

In this note, we re-evaluate limits on the ununified standard model in
light of current measurements of precision electroweak observables
both at the $Z$-pole and from low-energies. We perform a global fit to
all the data using the techniques of Burgess, {\it et. al.}
\ref\burlon{C.P. Burgess, et. al.,
hep-ph/9312291, {\it Phys. Rev.} {\bf D49} (1994) 6115.}. We show that recent
LEP data now place a lower
bound on the masses of the extra $W$ and $Z$ of order 2 TeV.

The second section of this note reviews the ununified standard model.
The third explains the linear approximation used to find the changes
in the electroweak observables relative to their standard model
values.  The last two sections discuss the global fit and the results.

\newsec{The Ununified Standard Model}

As described in ref. \oldunun , this model is based on the electroweak
gauge group  $SU(2)_q \times SU(2)_\l \times U(1)$.  Left-handed
quarks and leptons transform as doublets under $SU(2)_q$ and
$SU(2)_\l$, respectively; right-handed quarks and leptons transform as
singlets under both $SU(2)$ gauge groups. The $U(1)$ is the
hypercharge group of the standard model.~\foot{See \oldunun\ for
comments on the use of additional fermions to cancel the
$SU(2)_q^{\,2}\times U(1)$ and  $SU(2)_\l^{\,2} \times U(1)$
anomalies.} The gauge covariant derivative is  \eqn\der{\del^\mu +
i\gq \,T_q^a \,W_{qa}^\mu  + i\gl \,T_\l^a \,W_{\l a}^\mu +i\gp \,Y
X^\mu ,} where $T_q^a$ and $T_\l^a$, $a=1$~to~3, are the $SU(2)$
generators and $Y$ generates hypercharge.  The gauge couplings may be
written  \eqn\eone{
\gq={e\over \sp\st}\,,\quad
\gl={e\over \cp\st}\,,\quad
\gp={e\over \ct}\,, }
in terms of the usual weak mixing angle $\theta_W$ and a new mixing angle
$\phi$.

The electroweak gauge group
spontaneously breaks to $U(1)_{em}$ which is generated by
$Q= T_{3q}+T_{3\l}+Y.$  This symmetry breaking occurs
occurs when two scalar fields, $\Phi$
and $\Sigma=\sigma+i\vec\tau\cdot\vec\pi$,  transforming respectively as
$(1,2)_{1/2}$ and $(2,2)_0$ acquire the vacuum expectation values (vev's)
\eqn\phivev{\left\langle\Phi\right\rangle=\pmatrix{ 0\cr {v/{\sqrt
2}}\cr}\,,\quad\left\langle\Sigma\right\rangle=\pmatrix {u&0\cr
0&u\cr}\,\,\,.}
The vev of $\Sigma$ breaks the two \su2's down to the diagonal
\su2\ of the standard model.   Thus this theory reproduces the
phenomenology of the standard model for $u\gg v$.  What was originally
interesting about the model was that $u\approx v$ was permitted by
existing data for a wide range of $\sin\phi$.

In order to compare the model with present data, we will need to
understand both the form of the four-fermion current-current
interactions at zero momentum transfer and the properties of the
gauge boson eigenstates.  Let us start with the low-energy theory.
Putting the matrix of squared vev's in the $\pmatrix{q\cr \ell \cr}$
basis,
\eqn\vtwo{V^2=\pmatrix{u^2&-u^2\cr -u^2&v^2+u^2}}
and writing the left-handed
charged quark and lepton currents as $j_q^\mu$ and $j_\ell^\mu$, one
finds the charged current four-fermion weak interactions
\eqn\wcc{{2 \over
v^2}\left( j_q + j_\l \right)^2 + {2 \over u^2}j_q^2\,\,.}
Because the
non-leptonic weak interactions are enhanced by a factor $(1 +
v^2/u^2)$ relative to the leptonic and  semi-leptonic weak
interactions, the value of $v^2/u^2 \equiv 1/x$ must be less
than 1.  In studying this theory at energies below the weak scale, it
is therefore possible to use an effective theory corresponding to the
standard model plus corrections of order $1/x$.

Similarly, in terms of the neutral left-handed $T_3$ currents
$j^\mu_{3q}$ and $j^\mu_{3\ell}$ and the full electromagnetic current
$j^\mu_{em}$ the four-fermion neutral current interactions are
\eqn\ncplus{ {2\over
v^2}\,(j^\mu_{3q}+ j^\mu_{3\l} -\, j^\mu_{em}\sin^2\theta_W)^2 +{2\over
\,u^2}\, (j^\mu_{3q}-\, j^\mu_{em}\sin^2\phi\sin^2\theta_W)^2.}
Again, the first term has the same form as the neutral current
interactions of the standard model and the second term enhances
nonleptonic neutral currents.  What is different is that the second
term also contains new semileptonic and leptonic vectorial
interactions which vanish as $\sin^2\phi \to 0$.  Neutrino neutral currents
and the axial coupling of charged leptons are unaffected.

Next, we turn to the gauge boson eigenstates. It is convenient
\oldunun\ to rewrite the gauge bosons in the following  basis
\eqn\etwo{W^{\pm\,\mu}_1=s\,W^{\pm\,\mu}_q+c\,W^{\pm\,\mu}_\ell
\,,\quad  W^{\pm\,\mu}_2=c\,W^{\pm\,\mu}_q-s\,W^{\pm\,\mu}_\ell\,}
\eqn\ethr{Z_1^\mu=\ct\,(s\,W^\mu_{3q}+c\,W^\mu_{3\ell})-\st\,X^\mu
\,,\quad Z_2^\mu=c\,W^\mu_{3q}-s\,W^\mu_{3\ell}\,}
where $W_1$ and $Z_1$ are the standard model gauge bosons,
$s \equiv \sp$,
and $c \equiv \cp$.  Then in the limit that $1/x$ is  less
than 1, we can obtain perturbative expressions for the masses of
the light eigenstates
\eqn\mwzapprox{ {M^L_W\over M^0_W}
\approx  {M^L_Z\over M^0_Z}\approx
\left(1-{s^4\over 2\,x}\right)  ,}
where $M^0_W$ and  $M^0_Z$ are the
tree level gauge boson masses in the standard model. Note that, if
$s^2$ is small as well as $1/x$, the corrections to the masses are
small.   In the small $1/x$ limit
the light states are \eqn\wl{W^L\approx W_1+{s^3c\over
x}\,W_2\,,\quad  Z^L\approx Z_1+{s^3c\over x\,\cos\theta_W}\,Z_2\,\,.}
and they couple to fermions as, respectively,  \eqn\wlc{{e\over
\st}\left( T_q^\pm + T_\l^\pm  + {s^2 \over x} {\left(
 c^2 T_q^\pm - s^2 T_\l^\pm \right) } \right) }
\eqn\zlc{ {e \over {\st \ct}} \left( T_{3q} + T_{3\l} - \sin^2
 \theta_W \, Q +{s^2 \over x} {\left( c^2 T_{3q} - s^2 T_{3\l}
\right) } \right) \, .}
In this approximation, the heavy eigenstates have a mass given by
\eqn\mwzhapprox{ {M^H_W\over M^0_W}
\approx  {M^H_Z\over M^0_Z}\approx
{\sqrt{x}\over s c}\left(1+{s^4\over 2\,x}\right)\, .}

\newsec{Changes in Physical Observables}

At lowest-order, the predictions for electroweak observables in the
standard model depend only on the measured values of
$\alpha_{em}(M_Z)$, $G_F$, and $M_Z$. In the ununified model, the
lowest-order predictions will also depend on the values of $s^2$ and
$1/x$. To constrain $s^2$ and $1/x$, one
should fit the observed values of the precisely measured electroweak
quantities to their predicted values in the ununified model and
determine the allowed values of $s^2$ and $1/x$.

In practice we know that the standard model is at least
approximately correct and we expect that $1/x$ is small. Therefore, as
we have done in the previous section, we will calculate the values of
electroweak observables to leading order in $1/x$.  Using the
expressions given in the previous section, we may evaluate the changes
in various physical observables relative to their standard model
values to first-order in $1/x$ \burlon .  We obtain the following expressions
for these changes:

\eqn\dgamz{\Gamma_Z = \Gamma^{SM}_Z \left( 1 + {1\over x}\left[
0.732 s^4 + 1.634 s^2 c^2 \right] \right) ~,}

\eqn\drlep{R_\ell = {{\Gamma_h}\over{\Gamma_\ell}}=
R^{SM}_\ell \left( 1 + {1\over x}\left[
2.405 s^4 + 2.337 s^2 c^2 \right] \right) ~,}

\eqn\dsigh{\sigma_h =
{{12\pi \Gamma_e \Gamma_h}\over{M_Z^2\Gamma_Z^2}}=
 \sigma^{SM}_h \left( 1 + {1\over x}\left[
-0.931 s^4 - 0.931 s^2 c^2 \right] \right) ~,}

\eqn\drbb{R_b ={{\Gamma_b}\over{\Gamma_h}}=
 R^{SM}_b \left( 1 + {1\over x}\left[
-0.059 s^4 - 0.052 s^2 c^2 \right] \right) ~,}

\eqn\dalfb{A^\ell_{FB} = A^{\ell,SM}_{FB} + {1\over x}\left[
0.184 s^4 \right]~,}

\eqn\dabfb{A^b_{FB} = A^{b,SM}_{FB} + {1\over x}\left[
0.559 s^4 + 0.017 s^2 c^2 \right]~,}

\eqn\dacfb{A^c_{FB} = A^{c,SM}_{FB} + {1\over x}\left[
0.525 s^4 + 0.094 s^2 c^2 \right]~,}

\eqn\dalr{A_{LR} =A_e= A^{SM}_{LR} + {1\over x}\left[
0.769 s^4 \right]~,}

\eqn\datpt{A_\tau(P_\tau) = A^{SM}_{pol} (\tau) + {1\over x}\left[
0.769 s^4 \right]~,}

\eqn\daept{A_e (P_\tau) = A^{SM}_e (P_\tau) + {1\over x}\left[
0.769 s^4 \right]~,}

\eqn\dmw{M_W = M^{SM}_W \left( 1 + {1\over x}\left[
0.213 s^4 \right] \right)~, }

\eqn\dmw{{M_W \over M_Z} = {M^{SM}_W \over M^{SM}_Z} \left( 1 +
{1\over x}\left[ 0.213 s^4 \right] \right) ~,}

\eqn\dglsq{ g^2_L (\nu N \to \mu^- X) = (g^2_L)^{SM} + {1\over x}
\left[ 0.244 s^4 \right] ~,}

\eqn\dgrsq{ g^2_R (\nu N \to \mu^- X) = (g^2_R)^{SM} + {1\over x}
\left[ - 0.085 s^4 \right] ~,}

\eqn\dgev{ g_{eV}(\nu e \rightarrow \nu e) = g^{SM}_{eV} +
{1\over x}\left[ - 0.656 s^4 \right]~,}

\eqn\dgea{ g_{eA}(\nu e \rightarrow \nu e) = g^{SM}_{eA}~,}

\eqn\dqwcs{ Q_W(^{135}_{55} Cs) = Q_W^{SM} + {1\over x}\left[- 1.45 s^4
\right]\ .}
\noindent
Where
\eqn\dafb{A^f_{FB} = {{3}\over{4}} A_e A_f ~,}
\eqn\daf{A_f = {{2 g_V^f g_A^f}
\over{\left(g_V^f\right)^2+\left(g_A^f\right)^2}}~ ,}
\eqn\dgv{g_V^f = T_3 - 2Q \sin^2\theta_W~ ,}
and
\eqn\dga{ g_A^f=T_3~ .}
\noindent
Using the current experimental values of the electroweak observables
and using the corresponding best-fit {\it standard model} predictions,
we may use the equations above to fit the ununified model predictions
to the data.

\newsec{Global Fit}

Before proceeding with the fit and determining the allowed values of
$s^2$ and $1/x$, we must discuss the issue of higher-order
corrections.  At higher-order, the predictions of the standard or
ununified models also
depend\foot{The predictions also depend to a lesser extent
on the mass of the Higgs boson and, for the ununified model,
 the $\Phi$ and $\Sigma$
masses. At the
present level of experimental accuracy, this dependence is not
numerically significant.}  on the values of $\alpha_s(M_Z)$ and the
top-quark mass $m_t$. Given the success of the standard model, we
expect that, for the allowed range of $s^2$ and $1/x$, the changes in
the predicted values of physical observables due to radiative
corrections in the standard or ununified model will be approximately
the same {\it for the same values of $\alpha_s(M_Z)$ and $m_t$}.

The best-fit standard model predictions which we use
\ref\langacker{P. Langacker, hep-ph/9408310; see also
P. Langacker and J. Erler,
http://www-pdg.lbl.gov/www/rpp/book/page1304.html,
{\it Phys. Rev.} {\bf D50} (1994) 1304.} are based on a top quark
mass of 173 GeV (taken from a fit to precision electroweak data)
which, fortuitously, is consistent with the
range of masses ($174\pm16$ GeV) preferred by observed
top-candidate events at CDF
\ref\cdf{F. Abe et. al. (CDF collaboration), hep-ex/9405005,
{\it Phys. Rev. Lett} {\bf 73} (1994) 225; ibid.,
{\it Phys. Rev.} {\bf D50} (1994) 2966.}.

The treatment of $\alpha_s(M_Z)$ is more problematic: the LEP
 determination  for $\alpha_s(M_Z)$ comes
from a {\it fit}  to electroweak observables {\it assuming} the
validity of the standard model. For this reason, as emphasized by
Erler and Langacker
\ref\langackerbb{J. Erler and P. Langacker, hep-ph/9411203, UPR-0632-T.},
 when analyzing non-standard
models it is important to understand how the bounds vary for different
values of $\alpha_s(M_Z)$. We present results for bounds on $s^2$ and
$1/x$ both for $\alpha_s(M_Z) = 0.124$ (which is the LEP best-fit
value assuming the standard model is correct \langacker)
and for $\alpha_s(M_Z)=0.115$ as suggested by recent lattice
results \ref\lattice{C.T.H. Davies et. al., hep-ph/9408328,
OHSTPY-HEP-T-94-013, FSU-SCRI-94-79.} and deep-inelastic
scattering
\langacker\ref\deep{I. Hinchliffe,
http://www-pdg.lbl.gov/www/rpp/book/page1297.html,
{\it Phys. Rev.} {\bf D50} (1994) 1297;
M. Virchaux, Saclay preprint, DAPHNIA/SPP 92-30, presented
at the ``QCD 20 years later" workshop,  Aachen Germany, (1992);
R. Voss, in {\it Proceedings of the 1993 International Symposium on Lepton
 and Photon Interactions at High Energies}, Ithaca NY (1993).}.
 To the accuracy to which
we work, the $\alpha_s$ dependence of the standard model predictions
only appears in the $Z$ partial widths (we neglect the effect of the
uncertainty in $\alpha_s$ in the forward-backward asymmetries  since
this effect is small
compared to the experimental errors \ref\forback{A. Djouadi, B. Lampe, and
P.M. Zerwas, hep-ph/9411386, MPI-Ph/94-81, UdeM-GPP-Th-94-09,
DESY 94-201.}), and we use \langacker

\eqn\gammaalph{
\Gamma_q = \Gamma_q|_{\alpha_s=0}\left(1+ {{\alpha_s}\over{\pi}}
+ 1.409 \left({{\alpha_s}\over{\pi}}\right)^2 -
12.77\left({{\alpha_s}\over{\pi}}\right)^3 \right)}
to obtain the standard model predictions for $\alpha(M_Z)=0.115$.

We perform a global fit \burlon\ for the parameters of the ununified
model to all precision electroweak data: the $Z$ line shape, forward
backward asymmetries, $\tau$ polarization, and left-right asymmetry
measured at LEP and SLC; the $W$ mass measured at FNAL and
UA2; the
electron and neutrino neutral current couplings determined by
deep-inelastic scattering; and the degree of atomic parity violation
measured in Cesium.  Care was taken {\it not}
to use a Pentium based computer \ref\globe{A. Zitner, {\it The Boston Globe},
``IBM stops selling computers with flawed Intel chip", (Dec. 13,
1994) 1.}.  The experimental values \langacker
\ref\blondel{A. Blondel,
http://alephwww.cern.ch/ALEPHGENERAL/reports/reports.html,
CERN PPE/94-133.}
of the
electroweak observables used and the corresponding standard model
predictions \langacker \ are shown in Table 1.

\topinsert
$$
\vbox{\offinterlineskip
\hrule
\halign{&\vrule#&
\vrule\strut\quad\hfil#\hfil&
\quad\hfil#\hfil&
\vrule\quad\hfil#\quad\cr
& Quantity && Experiment && SM && Ununified &\cr
\noalign{\hrule}
&$\Gamma_Z$ && 2.4976 $\pm$ 0.0038 && 2.4923 && 2.4969  &\cr
&$R_e$ && 20.86 $\pm$ 0.07 && 20.731 && 20.807  &\cr
&$R_\mu$ && 20.82 $\pm$ 0.06 && 20.731 && 20.807  &\cr
&$R_\tau$ && 20.75 $\pm$ 0.07 && 20.731 && 20.807  &\cr
&$\sigma_h$ && 41.49 $\pm$ 0.11 && 41.50 && 41.44  &\cr
&$R_b$ && 0.2202 $\pm$ 0.0020 && 0.2155 && 0.2155  &\cr
&$A_{FB}^e$ && 0.0156 $\pm$ 0.0034 && 0.016 && 0.016  &\cr
&$A_{FB}^\mu$ && 0.0143 $\pm$ 0.0021 && 0.016 && 0.016  &\cr
&$A_{FB}^\tau$ && 0.0230 $\pm$ 0.0026 && 0.016 && 0.016  &\cr
&$A_{\tau}(P_\tau)$ && 0.143 $\pm$ 0.010 && 0.146 && 0.147  &\cr
&$A_{e}(P_\tau)$ && 0.135 $\pm$ 0.011 && 0.146 && 0.147  &\cr
&$A_{FB}^b$ && 0.0967 $\pm$ 0.0038 && 0.1026 && 0.1030  &\cr
&$A_{FB}^c$ && 0.0760 $\pm$ 0.0091 && 0.073 && 0.074  &\cr
&$A_{LR}$ && 0.1637 $\pm$ 0.0075 && 0.146 && 0.147  &\cr
&$M_W$ && 80.17 $\pm$ 0.18 && 80.34 && 80.35  &\cr
&$M_W/M_Z$ && 0.8813 $\pm$ 0.0041 && 0.8810 && 0.8811  &\cr
&$g_L^2(\nu N \rightarrow \nu X)$ && 0.3003 $\pm$ 0.0039 && 0.303 && 0.303
&\cr
&$g_R^2(\nu N \rightarrow \nu X)$ && 0.0323 $\pm$ 0.0033 && 0.030 && 0.030
&\cr
&$g_{eA}(\nu e \rightarrow \nu e)$ && -0.503 $\pm$ 0.018 && -0.506 && -0.506
&\cr
&$g_{eV}(\nu e \rightarrow \nu e)$ && -0.025 $\pm$ 0.019 && -0.039 && -0.040
&\cr
&$Q_W(^{135}_{55} Cs)$ && -71.04 $\pm$ 1.81 && -72.78 && -72.78  &\cr
}\hrule
}
$$
\medskip\noindent
{\bf Table 1}: Experimental \langacker\blondel \ and predicted values of
electroweak
observables for the standard model and ununified standard model
for $\alpha_s(M_Z)=0.115$ and (for the ununified model) $s^2=0.5$.
The standard model values correspond to the best-fit values (with
$m_t=173$ GeV, $m_{\rm Higgs} = 300$ GeV) in
\langacker, corrected for the change in $\alpha_s(M_Z)$, and the revised
extraction \ref\swartz{M. Swartz,  hep-ph/9411353, SLAC-PUB-6710; see also
A.D. Martin and D. Zeppenfeld, hep-ph/9411377, MADPH-94-855.}
of $\alpha_{em}(M_Z)$.
\bigskip
\endinsert

We present results of the fit in terms of limits on the mass of the
heavy $W$ gauge boson, $M^H_W$ (which is lighter than $M^H_Z$ by a
factor of $\cos\theta_W$), as a function  of the mixing
angle $s^2$. In figures
1 and 2 we show the 95\% (solid) and 90\% (dashed) confidence contours
in the $M^H_W$-$s^2$ plane for $\alpha_s(M_Z)=0.115$ and 0.124,
respectively. In both cases, we find that the lower bound on $M^H_W$
is approximately 2 TeV.

For $\alpha_s=0.115$ the standard model does not fit the data
particularly well.  The $\chi^2/{\rm df}$ for the standard model is
1.60, where the number of degrees of freedom (df) is the number of
measurements (21 since we are not assuming lepton universality) minus
the number of fit parameters (i.e. 0).  If the standard model were correct,
then there would be a 4\% probability that the fit would be this bad
or worse.

\topinsert
$$
\vbox{\offinterlineskip
\hrule
\halign{&\vrule#&
\vrule\strut\quad\hfil#\hfil&
\quad\hfil#\hfil&
\vrule\quad\hfil#\quad\cr
& Quantity && Experiment && SM && Ununified &\cr
\noalign{\hrule}
&$\Gamma_Z$ && 2.4976 $\pm$ 0.0038 && 2.4974 && 2.4980 &\cr
&$R_e$ && 20.86 $\pm$ 0.07 && 20.791 && 20.802  &\cr
&$R_\mu$ && 20.82 $\pm$ 0.06 && 20.791 && 20.802  &\cr
&$R_\tau$ && 20.75 $\pm$ 0.07 && 20.791 && 20.802  &\cr
&$\sigma_h$ && 41.49 $\pm$ 0.11 && 41.45 && 41.44  &\cr
&$R_b$ && 0.2202 $\pm$ 0.0020 && 0.2155 && 0.2155  &\cr
&$A_{FB}^e$ && 0.0156 $\pm$ 0.0034 && 0.016 && 0.016  &\cr
&$A_{FB}^\mu$ && 0.0143 $\pm$ 0.0021 && 0.016 && 0.016  &\cr
&$A_{FB}^\tau$ && 0.0230 $\pm$ 0.0026 && 0.016 && 0.016  &\cr
&$A_{\tau}(P_\tau)$ && 0.143 $\pm$ 0.010 && 0.146 && 0.146  &\cr
&$A_{e}(P_\tau)$ && 0.135 $\pm$ 0.011 && 0.146 && 0.146  &\cr
&$A_{FB}^b$ && 0.0967 $\pm$ 0.0038 && 0.1026 && 0.1027  &\cr
&$A_{FB}^c$ && 0.0760 $\pm$ 0.0091 && 0.073 && 0.073  &\cr
&$A_{LR}$ && 0.1637 $\pm$ 0.0075 && 0.146 && 0.146  &\cr
&$M_W$ && 80.17 $\pm$ 0.18 && 80.34 && 80.34  &\cr
&$M_W/M_Z$ && 0.8813 $\pm$ 0.0041 && 0.8810 && 0.8810  &\cr
&$g_L^2(\nu N \rightarrow \nu X)$ && 0.3003 $\pm$ 0.0039 && 0.303 && 0.303
&\cr
&$g_R^2(\nu N \rightarrow \nu X)$ && 0.0323 $\pm$ 0.0033 && 0.030 && 0.030
&\cr
&$g_{eA}(\nu e \rightarrow \nu e)$ && -0.503 $\pm$ 0.018 && -0.506 && -0.506
&\cr
&$g_{eV}(\nu e \rightarrow \nu e)$ && -0.025 $\pm$ 0.019 && -0.039 && -0.039
&\cr
&$Q_W(^{135}_{55} Cs)$ && -71.04 $\pm$ 1.81 && -72.78 && -72.78  &\cr
}\hrule
}
$$
\medskip\noindent
{\bf Table 2}: Experimental \langacker\blondel \  and predicted values of
electroweak
observables for the standard model and ununified standard model
for $\alpha_s(M_Z)=0.124$ and (for the ununified model) $s^2=0.5$.
The standard model values correspond to the best-fit values (with
$m_t=173$ GeV, $m_{\rm Higgs} = 300$ GeV)  in
\langacker, corrected for the revised
extraction \swartz \ of $\alpha_{em}(M_Z)$.
\bigskip
\endinsert

In contrast, for $\alpha_s=0.115$ in the ununified model (with
$s^2=0.5$) we find $\chi^2/{\rm df}=1.39$ with ${\rm df}=20$. If the
ununified  model were correct, the probability of a fit
this bad or worse would be 11\%, making the ununified model a
{\it better} fit to the data.   Furthermore, the standard model
actually lies outside of the 95\% confidence region surrounding the
best fit for $M^H_W$.
This results in the upper set of curves in the $M^H_W$-$s^2$
plane -- an upper bound on $M^H_W$ as a
function of the mixing angle. The best fit for the heavy $W$ mass
is $M^H_W= 2.9\,^{+0.9 }_{-0.5}{\rm TeV}$.

For $\alpha_s=0.124$ the standard model fit improves considerably.
The $\chi^2/{\rm df}$ for the standard model is 1.38 (here ${\rm
df}=20$ for both models, since $\alpha_s(M_Z)$ is a fit parameter for
the standard model), while for the ununified model (with $s^2 = 0.5$)
one also finds $\chi^2/{\rm df}=1.38$.  The probability of a fit with
$\chi^2/{\rm df}$ equal to or greater than that observed is 12\% for
both models.  The best fit
for the heavy $W$ mass at $s^2 = 0.5$ is
$M^H_W= 7.8\, ^{+\infty}_{-3.9}{\rm TeV}$.
The values of the electroweak observables used
and the corresponding model predictions are shown in Table 2.
\newsec{Discussion}

The ununified standard model provides a novel extension of the usual
$SU(2)\times U(1)$ gauge sector in which, at high-energies, leptons and
quarks transform under different weak $SU(2)$ gauge groups.  In this
note we have presented limits on the ununified standard model derived
from a global fit to all precision electroweak data. We find that the
model is now tightly constrained. In particular, at the 95\% confidence
level, the lower bound on the mass of the heavy $W$ and $Z$ is
approximately 2 TeV.

Heavy $W$ and $Z$ bosons weighing a few TeV should be visible at the
LHC in leptonic decay modes.  Since the heavy gauge bosons couple to
quarks with strength proportional to $c/s$ and to leptons as $s/c$,
the Drell-Yan cross-section (for small $s$) on the heavy boson resonance
would be of
order $(s/c)^4$ \oldunun .  The fact that the masses of the heavy $W$
and $Z$ are related by a factor of $\cos\theta_W$ \mwzhapprox\ would
help identify the gauge bosons as belonging to the ununified standard
model even if the bosons were so heavy or $s$ were so small that a
mere handful of events was observed.  The lower bound we have obtained
on the masses of the heavy $W$ and $Z$ implies that these bosons are
too massive to be produced at proposed electron-positron
colliders. Were a sufficiently energetic electron-positron machine to be
constructed, one would, correspondingly, expect the heavy $W$ and $Z$
to be visible in hadronic modes rather than leptonic ones.

\medskip
\centerline{\bf Acknowledgments}

We thank C. Burgess, J. Erler, and  P. Langacker for correspondence.
R.S.C. acknowledges the support of an Alfred P. Sloan Foundation
Fellowship, an NSF Presidential Young Investigator Award,
and a DOE Outstanding Junior Investigator Award.  EHS
acknowledges the support of an American Fellowship from the American
Association of University Women.
{\it This work was supported in part by the National Science
Foundation under grant PHY-9057173, by the Department of Energy under
grant DE-FG02-91ER40676.}

\listrefs

\centerline{\bf Figure Captions}
\bigskip\noindent
{\bf Figure 1} 90\% (dashed) and 95\% (solid) bounds on the mass of
the heavy $W$ gauge-boson of the ununified standard model ($M^H_W$)
as a function of $s^2$ for $\alpha_s(M_Z)=0.115$.  The allowed region
(at the specified confidence level) is between the curves.
\bigskip\noindent
{\bf Figure 2} 90\% (dashed) and 95\% (solid) bounds on the mass of
the heavy $W$ gauge-boson of the ununified standard model ($M^H_W$)
as a function of $s^2$ for $\alpha_s(M_Z)=0.124$.  The allowed region
(at the specified confidence level) is above the curve.

\vfill\eject

\epsfbox[120 210 420 660]{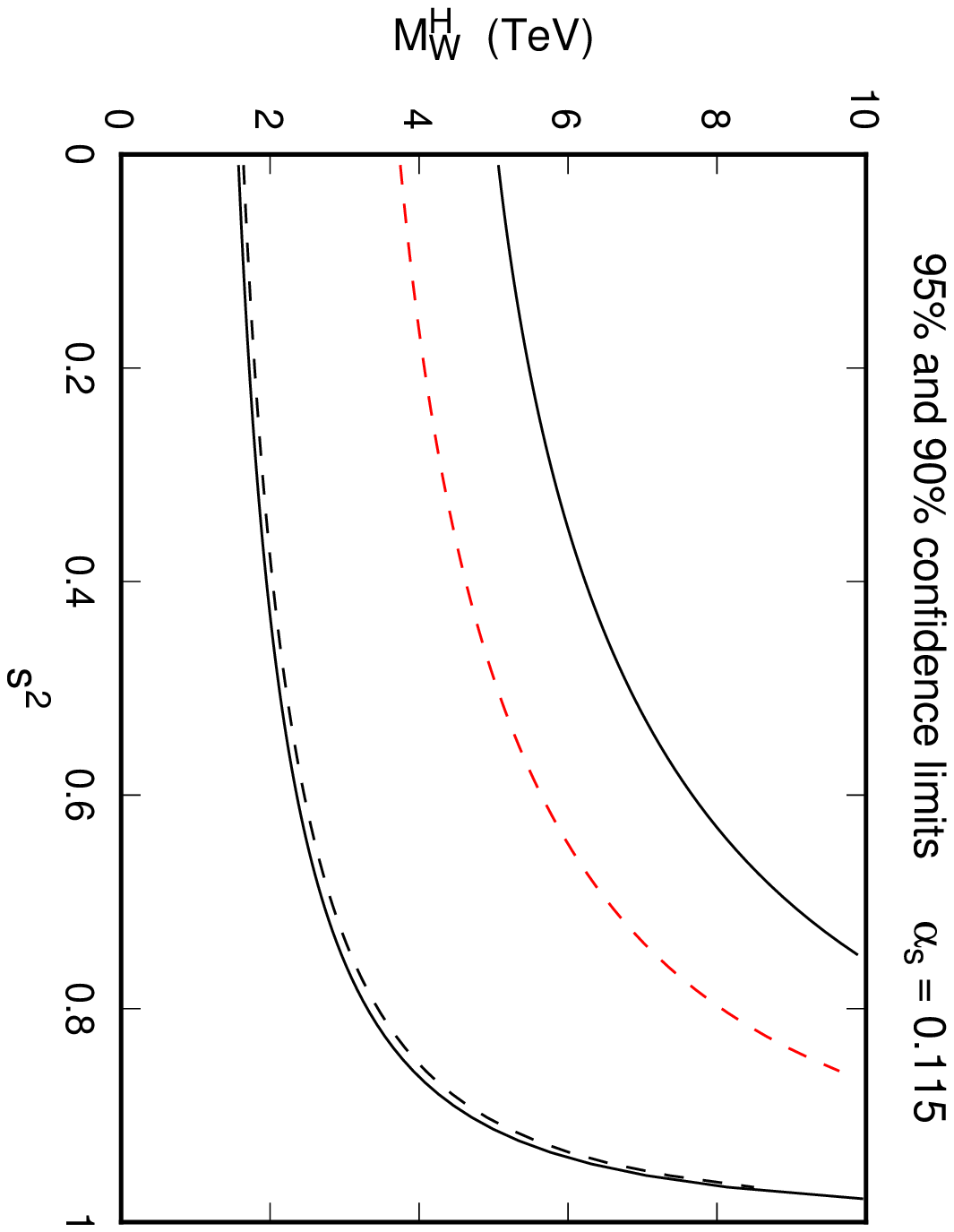}
\vfill\eject

\epsfbox[120 210 420 660]{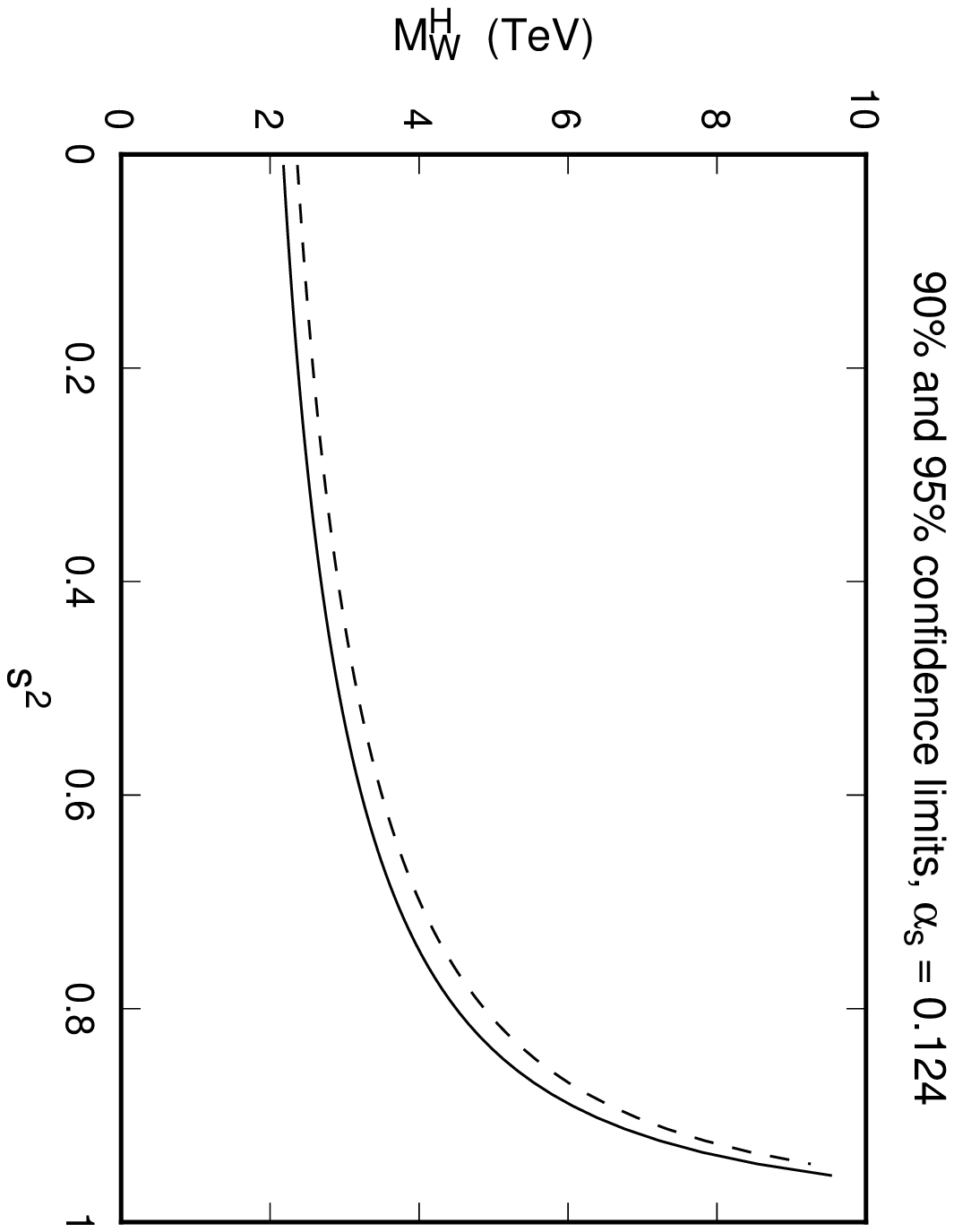}
\vfill\eject

\bye